\documentclass{article}
\usepackage{titlesec}
\usepackage{multicol}
\usepackage[english]{babel}
\usepackage[
backend=biber,
style=numeric,
citestyle=phys
]{biblatex}
\title{Bibliography management: \texttt{biblatex} package}
\usepackage[colorlinks=true, allcolors=blue]{hyperref}
\usepackage{physics}
\usepackage{indentfirst}
\usepackage[toc]{appendix}
\usepackage{graphicx} 
\usepackage[figurename=Figure]{caption}
\usepackage{float}    
\usepackage{verbatim} 
\usepackage{amsmath}  
\usepackage{amssymb}  
\usepackage{fullpage} 
\usepackage{paralist} 
\usepackage{listings} 
\usepackage{subfig}   
\usepackage{enumitem} 
\usepackage{siunitx}  

\usepackage{tikz,bm} 
\usepackage{circuitikz}
\usepackage{mathrsfs,amsmath} 
\usepackage [autostyle, english = american]{csquotes}
\MakeOuterQuote{"}
\usepackage{mathtools}

\usepackage[c]{esvect}
\usepackage{yfonts}
\usepackage{courier}
\renewcommand{\thesection}{\Roman{section}}
\titleformat{\section}{\large\bfseries}{\thesection.}{0.5em}{\MakeUppercase}
\usepackage{tocbibind}
\usepackage{glossaries}

\newglossaryentry{bit mask}
{
    name=bit mask,
    description={a technique used to encode and store additional information about each pixel in an image beyond its intensity value}
}

\newglossaryentry{normalize}
{
    name=normalize,
    description={adjusting the pixel values of the image to a common scale -- often the median value -- to enhance the visibility of certain features}
}

\newglossaryentry{zero point}
{
    name=zero point,
    description={calibration parameter that relates the instrumental magnitudes to a standardized magnitude system; it is essentially an offset or correction factor that brings the instrumental magnitudes onto a common scale}
}

\newglossaryentry{color coefficient}
{
    name=color coefficient,
    description={calibration parameter that describes the relationship between the difference in colors (which are, in turn, differences in brightness between two different passbands) of a source and its magnitude, quantifying how changes in the color of a source affect its measured magnitude}
}

\newglossaryentry{Vega-AB magnitude}
{
    name=Vega-AB magnitude,
    description={a standardized magnitude system used in astronomy to measure the brightness of celestial objects based on two key components: the Vega star (a bright star with a stable apparent magnitude often set to zero in all passbands) and the AB magnitude system (a standardized system that defines magnitudes based on the flux density of light)}
}

\newglossaryentry{passband}
{
    name=passband,
    description={a ``window" that selectively lets through certain colors of light while blocking or attenuating others}
}

\newglossaryentry{proper motion}
{
    name=proper motion,
    description={apparent motion of an object on the celestial sphere over time due to its actual motion through space. It is typically measured in units of angular displacement per unit time}
}

\newglossaryentry{Student's t-statistic}
{
    name=Student's t-statistic,
    description={a measure that helps assess whether the difference between two groups is statistically significant (or just due to random variation) by taking into account the average difference between the groups and the variability within each group}
}

\newglossaryentry{reduced chi2}
{
    name=reduced $\chi^2$,
    description={a measure of how well a given model fits the data, taking into account the number of data points (degrees of freedom) used in the analysis}
}

\newglossaryentry{adaptive threshold}
{
    name=adaptive threshold,
    description={a value that adjusts based on certain conditions to handle outliers and ensure that the probability is independent from number of trials; used for finding asteroids in the LAST proper motion search algorithm}
}

\newglossaryentry{false alarm probability}
{
    name=false alarm probability,
    description={likelihood of concluding that there is a significant effect in the data (e.g. a proper motion) when, in reality, there is no such effect}
}

\newglossaryentry{relative photometry}
{
    name=relative photometry,
    description={technique used to measure the magnitude of sources relative to each other by comparing their brightness in different epochs rather than obtaining absolute measurements of their brightness}
}

\newglossaryentry{light curve}
{
    name=light curve,
    description={plot of the flux from an astronomical source on the vertical axis against the corresponding time on the horizontal axis used to graphically represent how the source brightness changes over time}
}

\newglossaryentry{variability}
{
    name=variability,
    description={property of astronomical objects to exhibit changes in their brightness, luminosity, period, or other observable characteristics; can occur on different timescales, ranging from fractions of a second to billions of years}
}

\newglossaryentry{coadd}
{
    name=coadd,
    description={combine multiple successive sub-images of the same position into a single image after transforming each sub-image to the same reference frame, aligning them in a consistent and standardized manner}
}

\newglossaryentry{Register}
{
    name=register,
    description={align multiple images so that they are properly synchronized and spatially coherent by adjusting the images' position, rotation, and scale}
}

\newglossaryentry{World Coordinates System (WCS)}
{
    name=World Coordinates System (WCS),
    description={way to assign precise celestial coordinates (such as right ascension and declination) to each pixel in an image system used in astronomy to assign precise celestial coordinates to each pixel in an image}
}

\newglossaryentry{temporal cadence}
{
    name=temporal cadence,
    description={the frequency with which a telescope collects data over time; telescopes with high temporal cadence are designed to minimize the time interval between consecutive observations}
}

\newglossaryentry{weight}
{
    name=weight,
    description={numerical values associated with each pixel in an image used to represent the degree of contribution of each pixel when calculating the signal or detecting the source; used to account for the variations in light or counts caused by the PSF
}
}

\newglossaryentry{variable stars}
{
    name={variable stars},
    description={celestial objects whose brightness fluctuates over time due to intrinsic or extrinsic astrophysical phenomena}
}

\newglossaryentry{periodogram}
{
    name={periodogram},
    description={mathematical technique used to analyze time series data and identify periodic signals within the data by highlighting prominent frequencies}
}

\newglossaryentry{Large Array Survey Telescope (LAST)}
{
    name={Large Array Survey Telescope (LAST)},
    description={cost-effective multi-purpose astronomical observatory equipped with an array of telescopes designed to conduct wide-field imaging surveys of the sky in the visible-light spectrum}
}

\newglossaryentry{cost-effectiveness}
{
    name={cost-effectiveness},
    description={a measure of the efficiency of telescope resource allocation, defined as volume of space a telescope can probe per unit of time per unit cost}
}

\newglossaryentry{arc-second}
{
    name={arc-second},
    description={unit of angular measurement equal to 1/3600th of a degree, commonly used in astronomy to describe small angles on the celestial sphere}
}

\newglossaryentry{equatorial mounts}
{
    name={equatorial mounts},
    description={type of telescope mount used to align telescopes with the celestial coordinate system using e two axes, one aligned with the Earth's rotation axis and the other perpendicular to it}
}

\newglossaryentry{hour angle}
{
    name={hour angle},
    description={angular measurement used to express the westward rotation of a celestial object relative to the observer's meridian, typically measured in hours, minutes, and seconds to indicate the time elapsed since the object crossed the observer's meridian}
}

\newglossaryentry{declination}
{
    name={declination},
    description={celestial coordinate system analogous to latitude on Earth and measures the angle between the object and the celestial equator}
}

\newglossaryentry{visits}
{
    name={visits},
    description={process of observing a specific region of the sky for a predefined duration; multiple visits may be conducted for the same region to enhance data quality and ensure proper coverage}
}

\newglossaryentry{epochs}
{
    name={epochs},
    description={specific time periods when observations are made to capture variations over time}
}

\newglossaryentry{exposure time}
{
    name={exposure time},
    description={duration for which a telescope camera is exposed to light during an astronomical observation}
}

\newglossaryentry{field of view}
{
    name={field of view},
    description={angular extent (measured in degrees sqaure) of the sky that can be captured by a telescope camera (array), determining the area of the celestial sphere that can be observed simultaneously}
}

\newglossaryentry{bias image}
{
    name={bias image},
    description={image obtained by exposing a camera sensor to zero light, ensuring that only the electronic noise inherent in the detector is recorded}
}

\newglossaryentry{dark image}
{
    name={dark image},
    description={image taken with the camera sensor entirely covered and no light reaching it in order to capture the thermal noise or dark current of the detector arising due to the random generation of electrons within the sensor even without incoming light}
}

\newglossaryentry{flat image}
{
    name={flat image},
    description={image taken to correct for non-uniform pixel sensitivity and other optical imperfections in the telescope cameras, obtained by imaging a uniformly illuminated, featureless source, such as a dome or the twilight sky}
}

\newglossaryentry{sigma-clipped mean}
{
    name={sigma-clipped mean},
    description={statistical method used in data analysis to calculate the mean of a set of values while excluding outliers by iteratively removing data points that deviate significantly from the mean (beyond a specified number of standard deviations, known as the sigma threshold) and calculating the mean again}
}

\newglossaryentry{dark current}
{
    name={dark current},
    description={electrical signal generated within a camera sensor in the absence of incoming light introducing unwanted signal variations in the obtained images}
}

\newglossaryentry{potential well}
{
    name={potential well},
    description={region within each pixel where incoming photons generate electrons via the photoelectric effect}
}

\newglossaryentry{detector gain}
{
    name={detector gain},
    description={measure of the amplification efficiency of a detector by determining the number of electrons generated in the potential well per unit of incident photons}
}

\newglossaryentry{pixel overlap}
{
    name={pixel overlap},
    description={region where adjacent pixels in an image (or sub-image) share some common area due to the finite size of the pixels}
}

\newglossaryentry{near-edge effects}
{
    name={near-edge effects},
    description={distortions occurring at the edges of astronomical images due to the incomplete exposure of pixels near the image boundary leading to variations in photometric measurements near the image edges}
}

\newglossaryentry{Gaussian filter templates}
{
    name={Gaussian filter templates},
    description={mathematical functions used in matched-filtering techniques to convolve with astronomical images for signal enhancement; chosen for their ability to smooth the data while preserving spatial information, making them effective in detecting specific signals in images}
}

\newglossaryentry{pipeline}
{
    name={pipeline},
    description={series of automated and standardized data reduction and analysis steps applied to raw observational data, systematically calibrating, correcting, and extracting information from large volumes of data}
}

\newglossaryentry{mask image}
{
    name={mask image},
    description={binary image used to identify regions of an astronomical image that should be flagged during analysis, typically containing pixels with a value of 1 in regions of interest (e.g., sources or regions affected by near-edge effects) and 0 elsewhere}
}

\newglossaryentry{signal-to-noise ratio (SNR)}
{
    name={signal-to-noise ratio (SNR)},
    description={quantitative measure used to assess the quality of observational data by representing the ratio of the signal (e.g., the brightness of an astronomical source) to the background noise level}
}

\newglossaryentry{moment}
{
    name={moment},
    description={mathematical measure used to describe the shape, spread, and center of a distribution; the first moment is the mean, while the second moment yields the variance}
}

\newglossaryentry{photometry}
{
    name={photometry},
    description={process of quantifying the brightness or intensity of light emitted or reflected by astronomical sources by measuring the amount of light captured by a telescope or camera sensor and converting it into a numerical value}
}

\newglossaryentry{aperture photometry}
{
    name={aperture photometry},
    description={photometric method that involves measuring the total amount of light from an astronomical source within a predefined circular or elliptical aperture}
}

\newglossaryentry{Point Spread Function (PSF) fit photometry}
{
    name={Point Spread Function (PSF) fit photometry},
    description={photometric method that models the Point Spread Function (PSF) of an astronomical image to perform more accurate source measurements by convolving a theoretical PSF with the observed data and fits it to each detected source}
}

\newglossaryentry{GAIA-DR3 catalog}
{
    name={GAIA-DR3 catalog},
    description={third data release of the GAIA mission, a space-based astrometry mission by the European Space Agency (ESA), providing a comprehensive and precise astrometric and photometric database of over a billion stars, galaxies, and other celestial objects in the Milky Way and beyond}
}

\newglossaryentry{astrometric calibration}
{
    name={astrometric calibration},
    description={process of calibrating astronomical images to establish a precise correspondence between the pixel coordinates in the image and the corresponding celestial coordinates}
}

\newglossaryentry{astrometric solution}
{
    name={astrometric solution},
    description={set of transformation parameters that relate pixel coordinates in an astronomical image to their corresponding celestial coordinates (right ascension and declination) obtained through astrometric calibration}
}

\newglossaryentry{photometric calibration}
{
    name={photometric calibration},
    description={process of converting instrumental magnitudes obtained from astronomical observations into calibrated magnitudes on a standard photometric system by comparing the instrumental magnitudes of reference stars with known magnitudes from photometric standards}
}

\newglossaryentry{photometric error}
{
    name={photometric error},
    description={uncertainty or variation in the measured brightness of an astronomical source due to various factors, such as instrumental noise, atmospheric turbulence, and variations in the source's intrinsic brightness}
}

\newglossaryentry{interpolation}
{
    name={interpolation},
    description={mathematical technique used to estimate values between data points in a dataset to fill in missing data or to obtain continuous functions from discrete observations}
}

\newglossaryentry{variability estimators}
{
    name={variability estimators},
    description={statistical methods and algorithms used to quantify the degree and nature of variation in the brightness (or position) of astronomical sources over time}
}

\newglossaryentry{FITS}
{
    name={FITS},
    description={Flexible Image Transport System, a standard file format consisting of a header containing metadata and data arrays organized in a binary format}
}

\newglossaryentry{HDF5}
{
    name={HDF5},
    description={Hierarchical Data Format 5, a flexible and scalable file format providing a hierarchical structure that can store complex and multidimensional data, making it suitable for large and diverse datasets, such as astronomical images, simulations, and time-series data}
}

\newglossaryentry{sky surveys}
{
    name={sky surveys},
    description={systematic observations of the celestial sphere to map the distribution of celestial objects and probe astrophysical phenomena}
}

\newglossaryentry{pulsation}
{
    name={pulsation},
    description={periodic expansion and contraction of a star, causing its brightness to vary over time due to changes in surface area and, consequently, temperature and light emission}
}

\newglossaryentry{accretion}
{
    name={accretion},
    description={process of matter accumulating onto a celestial object, such as a star, planet, or black hole, due to gravitational attraction}
}

\newglossaryentry{rotation}
{
    name={rotation},
    description={spinning motion of a celestial object around its axis, revealing and concealing dark star-spots that cause fluctuations in brightness}
}

\newglossaryentry{eclipsing}
{
    name={eclipsing},
    description={phenomenon where one celestial object passes in front of another, leading to a temporary decrease in brightness as seen from an observer's perspective}
}

\newglossaryentry{raw image}
{
    name={raw image},
    description={unprocessed and uncalibrated digital representation of the data captured by a telescope camera containing the original pixel values recorded by the detector without any corrections for bias, dark current, flat-field, or other instrumental effects}
}

\newglossaryentry{calibration}
{
    name={calibration},
    description={process of adjusting raw data obtained from astronomical instruments to obtain accurate and scientifically meaningful measurements by applying various corrections.}
}

\newglossaryentry{CMOS}
{
    name={CMOS},
    description={Complementary Metal Oxide Semiconductor (CMOS), type of semiconductor technology used for photodetection, typically composed of multiple layers, with the light-sensitive part located at the top}
}

\newglossaryentry{matched-filter}
{
    name={matched-filter},
    description={signal processing technique used to enhance the detection and measurement of specific signals by convolving the image with a template or filter that designed to match the expected characteristics of the signal of interest}
}

\newglossaryentry{root mean square (RMS)}
{
    name={root mean square (RMS)},
    description={statistical measure used to quantify the average magnitude of a set of values or a signal calculated by taking the square root of the mean of the squared values}
}

\makeglossaries

\title{Searching for Variable Stars Using the Large Array Survey Telescope (LAST)}

\author{Barkotel F. Zemenu\\[1ex]
\small \textit{Department of Physics, Yale College, New Haven, CT 06511, USA} \\[0.0005ex]
\small \textit{Weizmann Institute of Science, Department of Particle Physics and Astrophysics, 76100 Rehovot, Israel\footnote{Kupcinet-Getz International Summer School 2023}}}
\begin{document}
\maketitle
\begin{abstract}
This paper introduces a novel variability report generator developed for the Large Array Survey Telescope (LAST), a cost-effective multi-purpose telescope array conducting a wide survey of the variable sky in the visible-light spectrum. Designed to automate variability detection, the report generator identifies candidate variable stars by employing adjustable thresholds to detect periodic and non-periodic variables. The program outputs a visual and tabular photometric report for each candidate variable source from a given LAST sub-image. Functioning as a whitepaper, this document also provides a concise overview of LAST, discussing its design, data workflow, and variability search performance.
\end{abstract}

\newpage 

\tableofcontents

\newpage

\section{Introduction} \Gls{sky surveys} are abundant yet imperfect -- from probing unexplored regions of the sky to conducting surveys more efficiently, there exist several areas of improvement. The science reach of such surveys is manifold, ranging from the discovery of exoplanets to follow up studies of gravitational waves to investigation of \gls{variable stars}. As the underlying astrophysics of the latter -- celestial objects with fluctuating brightness -- spans both intrinsic (eg. \gls{pulsation},  \gls{accretion}) and extrinsic (eg. \gls{rotation},  \gls{eclipsing}) phenomena, investigating variable stars carries considerable scientific interest. While large-scale sky surveys return a wealth of information about \gls{variability} in the sky, their infrastructure cost remains an exceedingly prohibitive factor in the effort to improve upon performances of existing sky surveys. The \gls{Large Array Survey Telescope (LAST)} is designed to address this issue. Relying on a system of multiple small telescopes constructed with readily available components -- instead of a single large custom-designed equivalent -- LAST boasts orders of magnitude improvement in \gls{cost-effectiveness} compared to existing sky survey systems. Operating in southern Israel, LAST's geographic location not only takes advantage of the Negev desert's clear night sky, but also allows for a probe of sky regions not covered by prior wide-field surveys.

\begin{figure*}[hbt!]
\centerline{\includegraphics[width=10cm]{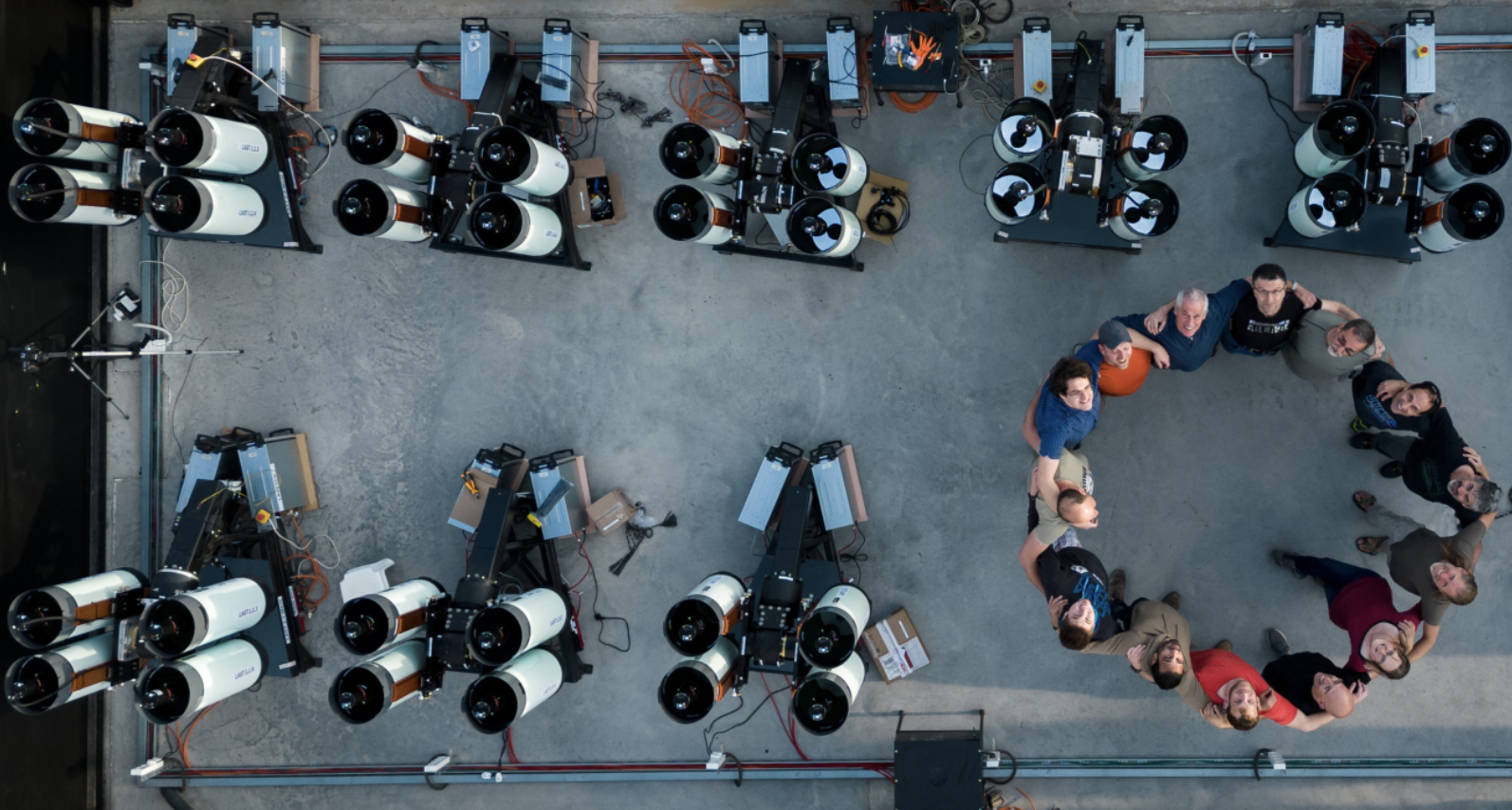}}
\caption{Top view of 32 installed LAST telescopes at Neot-Semadar, Israel} 
\label{topview}
\end{figure*}

This paper provides an overview of LAST's variability search performance. In Section \ref{lastobs}, the technical aspects of the LAST observatory describing the imaging strategy and system design are discussed. Section \ref{dataredpipe} details the data reduction pipeline, a crucial component in processing the vast amounts of data generated by LAST. The pipeline's calibration and processing steps are further outlined. In Section \ref{varsearch}, some techniques employed in variability searches are explored, along with their significance in identifying periodic and non-periodic variable stars. In Section \ref{varrefgen}, the variability report generator is presented, a novel custom-built code that automates the identification and reporting of candidate variable sources, streamlining the analysis process. Section \ref{conclusion} concludes with a brief summary, followed by an appendix and a dedicated glossary provided to ensure accessibility for novice researchers consulting advanced technical papers on LAST, such as those on its science goals \cite{sciencepaper}, {system overview} \cite{overviewpaper} and data reduction pipeline \cite{pipelinepaper}. 

\section{LAST Observatory} \label{lastobs} The LAST observatory consists of 48 telescopes (32 of which have been installed by summer 2023; see Fig. \ref{topview}), each equipped with affordable \gls{CMOS} detectors that allow for high \gls{temporal cadence} with an image quality in the order of an \gls{arc-second} and an image resolution roughly better than five iPhone 13 Pro cameras.\footnote{61 Mpix (LAST camera) vs. 12 Mpix (iPhone 13 Pro)} The telescopes are installed on robot-like structures -- better known as \gls{equatorial mounts} -- which allow for smooth telescope motion. Designed for precise tracking of celestial objects in the sky, each {mount} hosts four LAST telescopes that capture either a contiguous field of view or point to the same position in the sky. While two computers serve as the brain of each mount, two powerful motors assume the role of a muscle, one controlling telescope movement along the \gls{hour angle}, akin to moving the telescope left and right, while the other controls the movement of the telescope along the \gls{declination}, moving the telescope up and down. The motors are equipped with sensors that constantly monitor the telescope's position, allowing the mount to know exactly where the telescope is pointing at any given time.

\begin{figure*}[hbt!]
\centerline{\includegraphics[width=5cm]{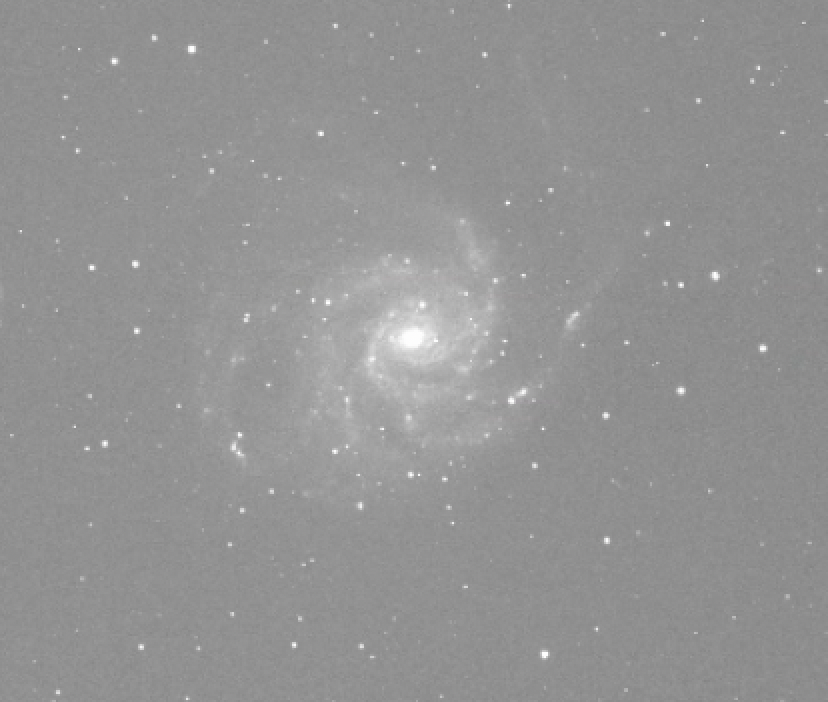}}
\caption{LAST raw image of a spiral galaxy} 
\label{rawimage}
\end{figure*}

\section{Data Reduction Pipeline}
\label{dataredpipe}
LAST operates using a "20 x 20" imaging strategy whereby a telescope camera captures 20 successive images (called \gls{epochs}) -- each with 20 seconds \gls{exposure time} -- with multiple \gls{visits} across a designated \gls{field of view}. The implementation of this strategy not only allows for the detection of phenomena with high temporal variability in short time frames but also for the study of dynamic celestial events with detectable changes in brightness. While scientifically advantageous, however, this imaging strategy generates an astounding amount of data at a staggering rate of 2.2 Gigabits per second (GBits/s).\footnote{This is akin to transferring the contents of over a 1000 phone pictures ($\sim 2$MBits) every second!} The enormously high data rate presents unique challenges, necessitating a robust data reduction \gls{pipeline} that efficiently handles tasks ranging from \gls{raw image} processing to processed imaged \gls{calibration} to variability detection. 

\subsection{Preliminary Image Calibration}
The initial step in the LAST pipeline involves a series of calibration procedures to ensure the data's reliability. We outline key components of the initial calibration in this section.

\subsubsection*{Bias Image Subtraction}
The \gls{bias image} accounts for the fluctuating noise of photon counts in the raw images. This image is captured with zero-exposure time and a closed shutter, effectively measuring the baseline noise inherent in the detector. By incorporating the bias image, we mitigate negative values that may arise due to noise during the image readout, ensuring a more stable data foundation.

\subsubsection*{Dark Image Subtraction}
The \gls{dark image} is taken by covering the camera lens with an opaque material (or with closed shutter) in order to determine the light produced by the camera itself. As a result, the dark image represents the combination of the baseline bias level and the dark current generated by the detector in the absence of light. Subtracting the dark image allows us to correct for both the bias level and the impact of dark current. The dark image is obtained by the "20 $\times$ 20" strategy every few weeks. The 20 images are combined into a "master" dark image via a \gls{sigma-clipped mean} in order to reject outlier pixels and suppress the effect of random noise.

\subsubsection*{Flat Image Subtraction}
Completing the calibration trio, the \gls{flat image} compensates for the non-uniformity of light on the camera's sensor. This calibration image is captured during twilight, a time when the light is assumed to be evenly distributed across the camera's field of view. Through normalizing the raw image by the flat image, variations due to non-uniform illumination in individual pixel sensitivities are accounted for, ensuring that each pixel's response is calibrated to represent the true intensity of the incoming light. Just like the dark image, the flat image is also obtained via the "20 $\cross$ 20" strategy. The individual images are normalized by their own image median before the combined flat image is yet again normalized by its own image median. The median pixel -- now equal to one -- would then serve as a reference point for determining the distribution of pixel values.

\begin{figure*}[hbt!]
\centerline{\includegraphics[width=5cm]{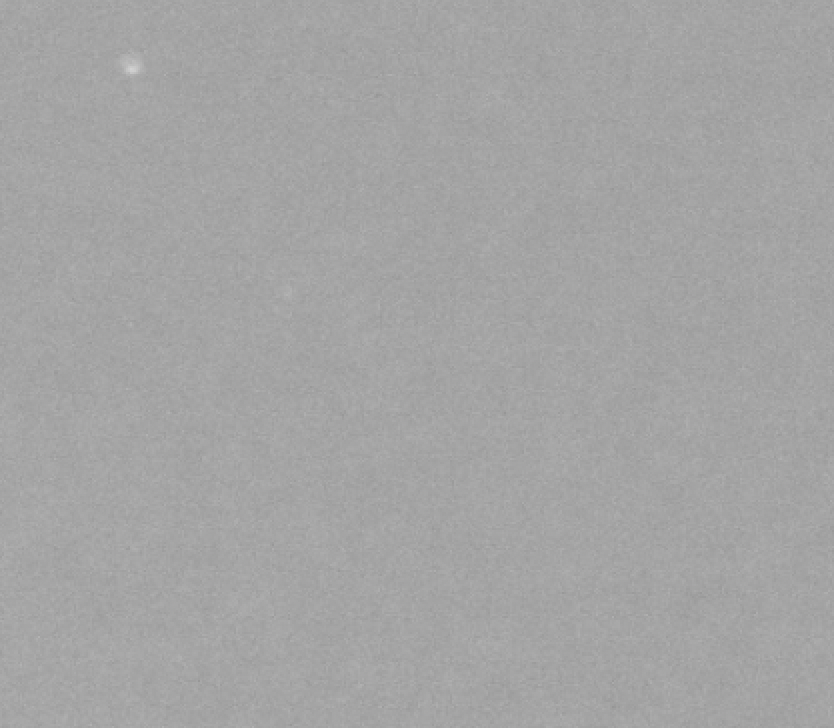}}
\caption{Flat image taken near twilight} 
\label{lastpic}
\end{figure*}
\subsubsection*{Bit Mask Representation}
To flag various pixel properties in the raw image, a 32-bit \gls{mask image} is generated. Based on their bit positions, the mask image can identify 32 different pixel states. Each bit position corresponds to a specific pixel attribute that is populated by propagation throughout the pipeline process. For instance, the \texttt{Saturation} bit ("Is pixel saturated?"), the \texttt{DarkHighVal} bit ("Is dark value too high?")  and the \texttt{FlatLowVal} bit ("Is normalized flat value too low?") are populated by the raw image, dark image and flat image respectively.

\subsubsection*{Additional Calibrations}
We perform additional calibration steps to further refine the raw image data. Chief among these is the issue of \gls{dark current}, which refers to the random electrical signal present in the detector even when no light is incident on it. This dark current can introduce unwanted noise into the image, potentially affecting the quality of our data. To mitigate this, we subtract the global median value of the darkened pixels, effectively removing the residual dark current and normalizing the dark signal across the entire image. We also tackle the non-linear response of the detector's \gls{potential well}, the region where the charges generated by incoming photons are collected and measured. Due to the non-linearity of this process, the recorded digital values may not precisely correspond to the actual number of photons detected. To account for this, we apply a correction that ensures the brightness levels in the calibrated image accurately represent the true intensity of the incoming light. Finally, we address the \gls{detector gain}, which refers to the conversion factor between the number of incident photons and the recorded digital values. The gain varies across different areas of the detector and can introduce non-uniformity in the image's intensity. By multiplying the pixel values with the appropriate detector gain factor, we achieve a uniform intensity level in the calibrated image. This step also compensates for any variations or non-linear effects introduced during the calibration process, ensuring the final image accurately reflects the original signal with the correct amplitude.


\subsection{Sub-image Processing}
Following preliminary image calibration, the LAST pipeline follows the procedure below for image processing:

\begin{enumerate}
    \item Partition the resulting image into 24 sub-images, accounting for \gls{pixel overlap} and \gls{near-edge effects}
    \begin{itemize}
        \item The reduction of the field of view increases calibration accuracy, improves performance speed and renders the file size more manageable
    \end{itemize}

    \item Subtract background from each sub-image, where background could emanate from the atmosphere, light pollution, etc
    \begin{itemize}
        \item Background is characterized by the \textit{mode} pixel value as it typically refers to the relatively homogeneous and uniform intensity level representing the underlying scene

        \item Applying linear \gls{interpolation}, map the total background and subtract it off from each sub-image
    \end{itemize}

    \item Generate a \gls{matched-filter} response (see Appendix \ref{app1}) by convolving each sub-image with five \gls{Gaussian filter templates} of varying widths to enhance signals well-correlated with the templates
    \begin{itemize}
        \item Normalize each matched-filter image by its {standard deviation (StD)} such that the resulting image is in units of StD
        
        \item Generate a maximized-pixel image by combining the highest value for each pixel across all five matched-filter images

        \item Search for source candidates by imposing a 5$\sigma$ threshold on local maxima in the combined matched-filter image
    \end{itemize}

    \item Create a catalog for each sub-image by identifying candidate sources and calculating their \gls{signal-to-noise ratio (SNR)}\footnote{LAST requires both the {detection} SNR (used for detecting sources; disregards noise contribution of source) and the {measurement} SNR (used for measuring source properties; takes into account noise arising from source)} to provide a measure of the detection significance

    \item Determine the average position (first \gls{moment}) and moment of inertia (second moment) of each source 
    \begin{itemize}
        \item Moments are calculated based on pixel intensities within each sub-image using a list of each source candidate's integer-pixel position
        
        \item The first moment provides the two-dimensional Cartesian coordinates that indicate the center location of the object while the second moment is related to the object's size, shape, orientation and elongation

        \item Moments are calculated iteratively, beginning with a flat \gls{weight} function (i.e. each pixel's contribution considered equally regardless of  intensity) until narrowing down to a pre-defined Gaussian width (i.e. the weight assigned to each pixel becomes a normal distribution)
    \end{itemize}

    \item Reject sources that resemble cosmic rays, i.e. with a delta function signature
    \begin{itemize}
        \item These pixels are identified using hypothesis testing to delta function and flagged in the bit mask image as cosmic rays
    \end{itemize}

    \item Perform \gls{photometry} (see Appendix \ref{app2})
    \begin{itemize}
        \item To work in flux units, normalize the filtered image and determine each source's flux value at its whole pixel position (e.g. $(3.2, 6.7) \to (3, 6)$)

        \item Perform \gls{aperture photometry} and \gls{Point Spread Function (PSF) fit photometry} to determine the magnitude of each source 
        
    \end{itemize}

    \item Perform \gls{astrometric calibration}
    \begin{itemize}
        \item Identify the precise celestial coordinates (viz., right ascension and declination) for each remaining stellar source (i.e. with a PSF -- and not delta function -- signature) by aligning each sub-image with the \gls{GAIA-DR3 catalog} reference frame
        
        \item Determine the \gls{astrometric solution} by matching star patterns in an image catalog to a reference catalog, without prior knowledge on the image’s sky position, rotation, or scale\footnote{Suffice to remark that the pattern matching process entails a two-step fitting procedure with an affine transformation and a 3rd order polynomial.}
    \end{itemize}

    \item Perform \gls{photometric calibration}
    \begin{itemize}
        \item Match the sources within each sub-image to corresponding sources in the GAIA-DR3 catalog
        
        \item Set a limit on the \gls{photometric error} and SNR to ensure that the matched sources have accurate measurements and are not affected by significant noise

        \item Fit for photometric parameters (namely the {\gls{zero point}} and {\gls{color coefficient}} using GAIA's {\gls{Vega-AB magnitude}} and {\gls{passband}})

        \item Extend the photometric calibration to all instrumental magnitudes so that the measured brightness values in the sub-images are consistent with the GAIA catalog\footnote{At this step, the photometry procedure also performs interpolation over pixels that are either saturated, dominated by dark current or numerically invalid (negative or NaN) using {local convolution} to obtain cleaner images} 
    \end{itemize}
    
    \item \label{step10}Match all sources in the 20 epochs of a visit (in a sub-image) to produce a visit-matched source catalogue
    \begin{itemize}
        \item Populate the first image with new sources that appear in any one of the 19 other epochs

        \item Match each source in the first image against all the sources in the other 19 images to generate a matrix -- {for every measured property}\footnote{These could be astrometric measurements (such as the declination or first-moment), photometric measurements (such as the PSF magnitude or background flux measurement), or bit-mask flag information} -- with 20 rows for the epochs and however-many columns for the sources

        \item Generate a merged catalog table using information from all 20 epochs, where each source populates a row and every relevant source information populates a column
    \end{itemize}

    \item \label{step11} Populate the merged catalog table with \gls{proper motion} fits and {\gls{variability estimators}}
    \begin{itemize}
        \item Fit source positions both to a no-proper-motion and linear-proper-motion model to statistically investigate\footnote{The statistical investigation involves, among others, a \gls{reduced chi2} for the null hypothesis, the \gls{Student's t-statistic}, an \gls{adaptive threshold} and a \gls{false alarm probability}} positional variation of sources, and identify the moving sources 

        \item Apply a \gls{relative photometry} zero point using the first image as a reference frame, i.e. align the magnitude system across all images by anchoring it to the photometric calibration of the first image

        \item Search for variable sources using variability estimators across the 20 epochs\footnote{This step is described in further detail in Section \ref{varsearch}} 
    \end{itemize}
    
    \item Cross-match each source with nearly two dozen external catalogs of sources 
    \begin{itemize}
        \item Identify which external catalog with which each source was matched using a bit integer mask
    \end{itemize}
    
    \item \Gls{coadd} the 20 epochs, refine the photometry and astrometry, and save the coadded catalog
    \begin{itemize}
        \item \Gls{Register} the epochs in each sub-image using the \gls{World Coordinates System (WCS)} to ensure the images are properly aligned with respect to their celestial coordinates

        \item Refine the photometry by summing the pixel values of corresponding pixels in the registered images, coadding using a {5$\sigma$ clipping} method, and estimating the background and variance in the coadd image

        \item Refine the astrometry by improving the positional information accuracy of the identified sources in the coadd image
    \end{itemize}

    \item Analyze the outputs from this pipeline process
    \begin{itemize}
        \item A single visit from each telescope's camera produces (1) a sky image data, (2) an associated bit mask image, and (3) a catalog of detected sources; these three data products are saved for each of the 24 sub-images obtained from each of the 20 epochs, resulting in a total of 1440 data files 
        
        \item The sources in each sub-image are then matched across epochs to produce a total of 24 merged catalog tables (in \gls{FITS} file format) and 24 merged catalog matrices (in \gls{HDF5} file format) as described in \ref{step10} and \ref{step11}

        \item Finally, each coadded sub-image of a visit produces (1) a data image, (2) a bit mask, (3) source catalog, and (4) PSF-fit photometry catalog (obtained via the refined photometry), resulting in a total of 96 files from a single visit using one telescope
    \end{itemize}
\end{enumerate}

\section{Variability Search}
\label{varsearch}
The calibration and processing steps described in \ref{dataredpipe} are intended to make raw data collected by LAST telescopes ripe for LAST's science goals.\footnote{LAST's science goals are thoroughly discussed in \cite{sciencepaper}.} Sifting through data collected over several visits, each with 20 epochs and 24 sub-images, we seek to detect and study variable sources. We describe below some of the tools we employ in this variability search.

\subsection{Light Curve}
The \gls{light curve} -- a plot of source brightness as a function of time -- visually represents the changes in a star's brightness over time, enabling the survey to identify recurring patterns or irregularities of interest. For instance, a pulsating star
will show periodic increases and decreases in brightness, creating a characteristic light curve. To enhance interpretability of the light curve, we employ two key techniques: binning and folding. 

\subsubsection*{Binning}
When we observe a star over time, we collect numerous data points representing its brightness at different moments. These data points can sometimes exhibit random fluctuations, making it challenging to discern any periodic variation. Binning involves grouping these individual data points into evenly spaced bins and then calculating an average value for each bin. By doing so, we reduce the noise in the data, resulting in a smoother representation of the star's brightness variations over time. This smoothed version of the light curve makes it easier to identify trends, patterns, and periodic behaviors that may be present in the data.

\begin{figure}[hbt!]
\centerline{\includegraphics[width=10cm]{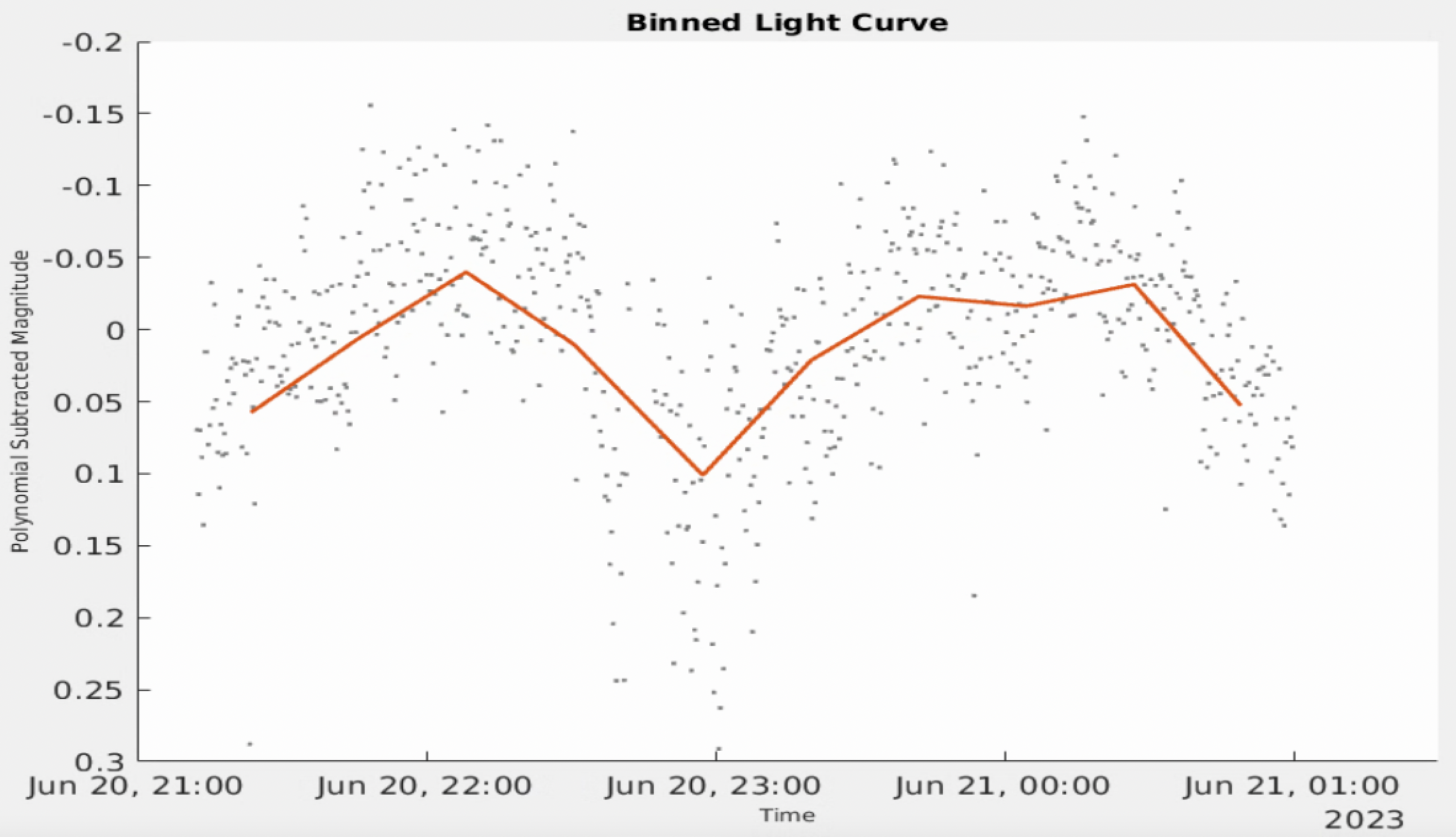}}
\caption{Binned Light curve} 
\label{}
\end{figure}

\subsubsection*{Folding}
When a star exhibits periodic behavior, its brightness variations repeat at regular intervals, forming a repetitive pattern. Folding aligns multiple cycles of the light curve on top of each other based on the star's period as determined from the dominant frequency in the periodogram. The process of folding involves taking each data point in the light curve and placing it at its corresponding phase within one period. The phase represents where each data point occurs in the repeating cycle of the star's brightness variation, normalized to a range of 0 to 1. By aligning the data points in this way, the folded light curve highlights the repeating pattern more clearly.

\begin{figure}[hbt!]
\centerline{\includegraphics[width=10cm]{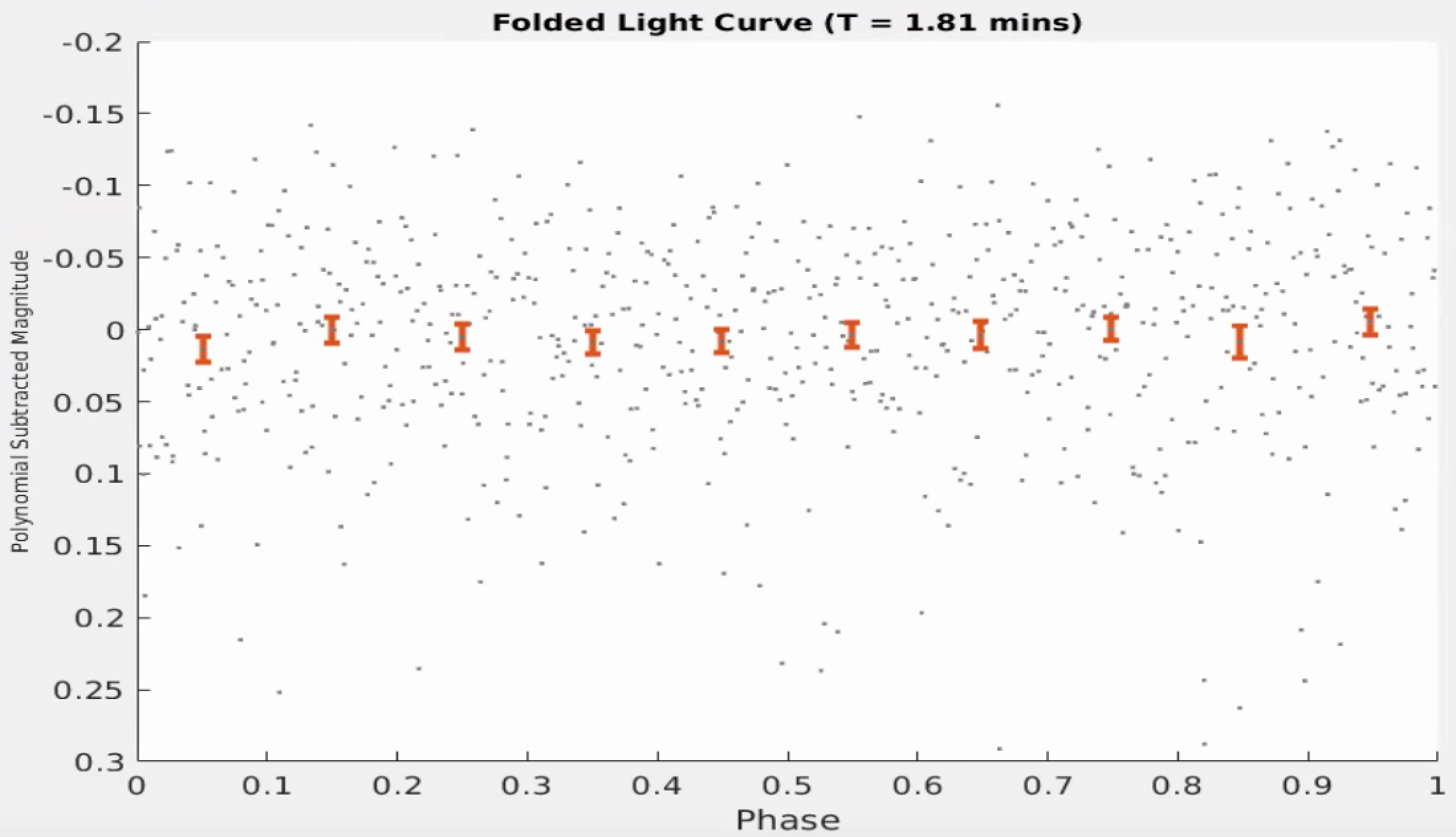}}
\caption{Folded Light curve} 
\label{}
\end{figure}

\subsection{Periodogram}
The \gls{periodogram} ("period" + "diagram") identifies dominant frequencies/periods associated with the periodic variations of a source. By converting time-domain data -- the light curve -- to the frequency domain, the periodogram represents the dataset as a sum of sinusoidal functions with varying frequencies. The resulting periodogram plot displays the distribution of power across different frequencies. While frequencies run along the x-axis of the plot, the y-axis of the periodogram represents the significance level of the peaks, measured in sigma, which indicates the statistical confidence of the identified frequencies. Peaks observed in the periodogram correspond to dominant frequencies associated with the periodic variations in the star's brightness. Analyzing these peaks enables us to accurately determine the period of variation for the star, which often reveals information about the star's physical properties, including its size, mass, and evolutionary stage. 

\begin{figure}[hbt!]
\centerline{\includegraphics[width=10cm]{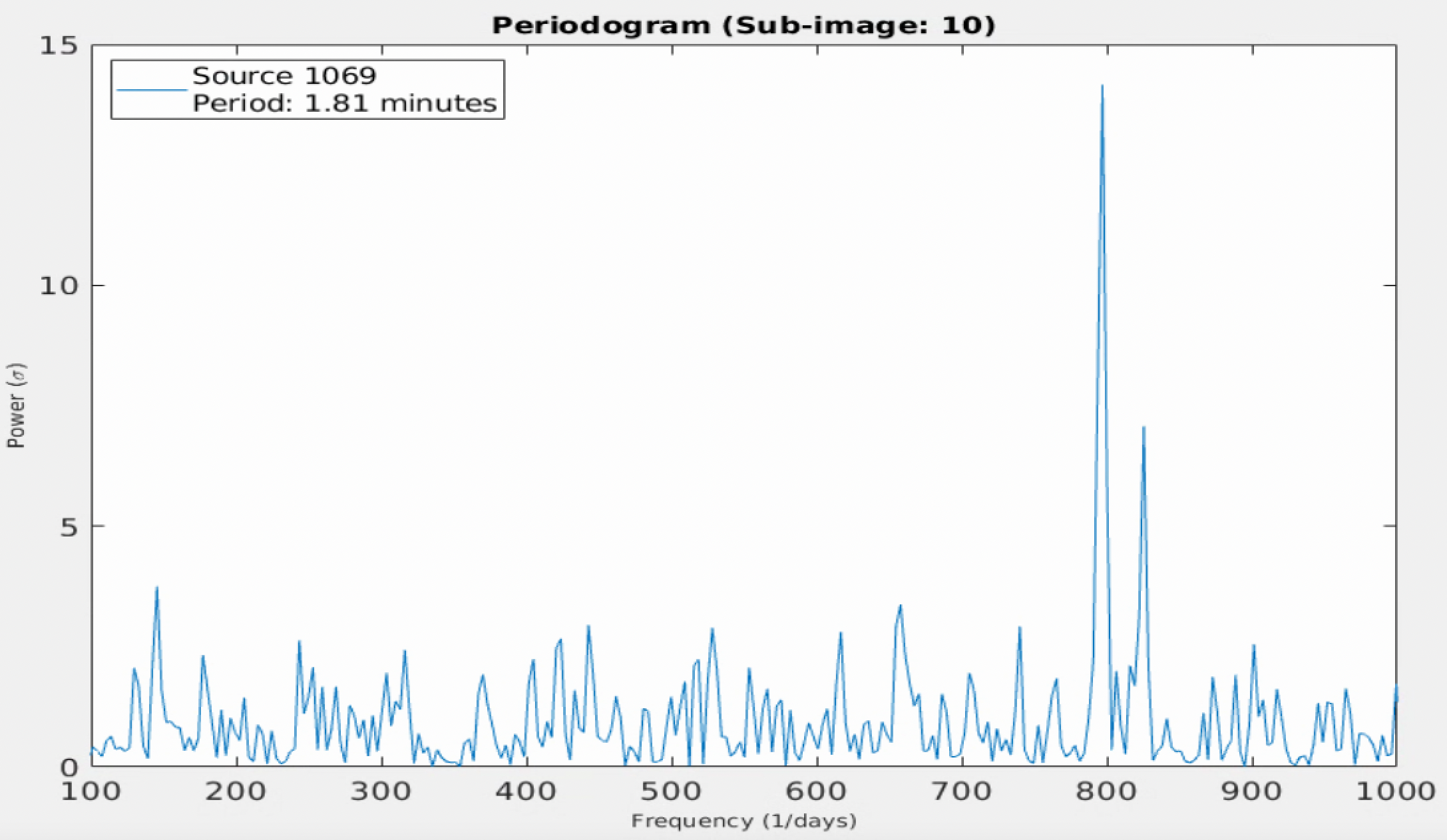}}
\caption{Periodogram (Amplitude vs. Frequency)} 
\label{}
\end{figure}

\subsection{plotRMS}
The \gls{root mean square (RMS)} plot (abbreviated plotRMS) provides a measure of the scatter of brightness measurements around the mean value for each star by calculating the RMS of the differences between the observed brightness and the average brightness over all the epochs in all visits. The plotRMS is particularly useful for identifying variable stars that exhibit irregular variability -- in contrast to stars with periodic variations -- as the periodogram will not identify them. Variable stars with larger RMS values indicate greater fluctuations in brightness, indicating a higher level of variability over time. Furthermore, The plotRMS assists in distinguishing true variable sources from potential noise or systematic errors. It allows us to set appropriate thresholds for variability detection and select significant variable candidates for further analysis.

\begin{figure}[hbt!]
\centerline{\includegraphics[width=10cm]{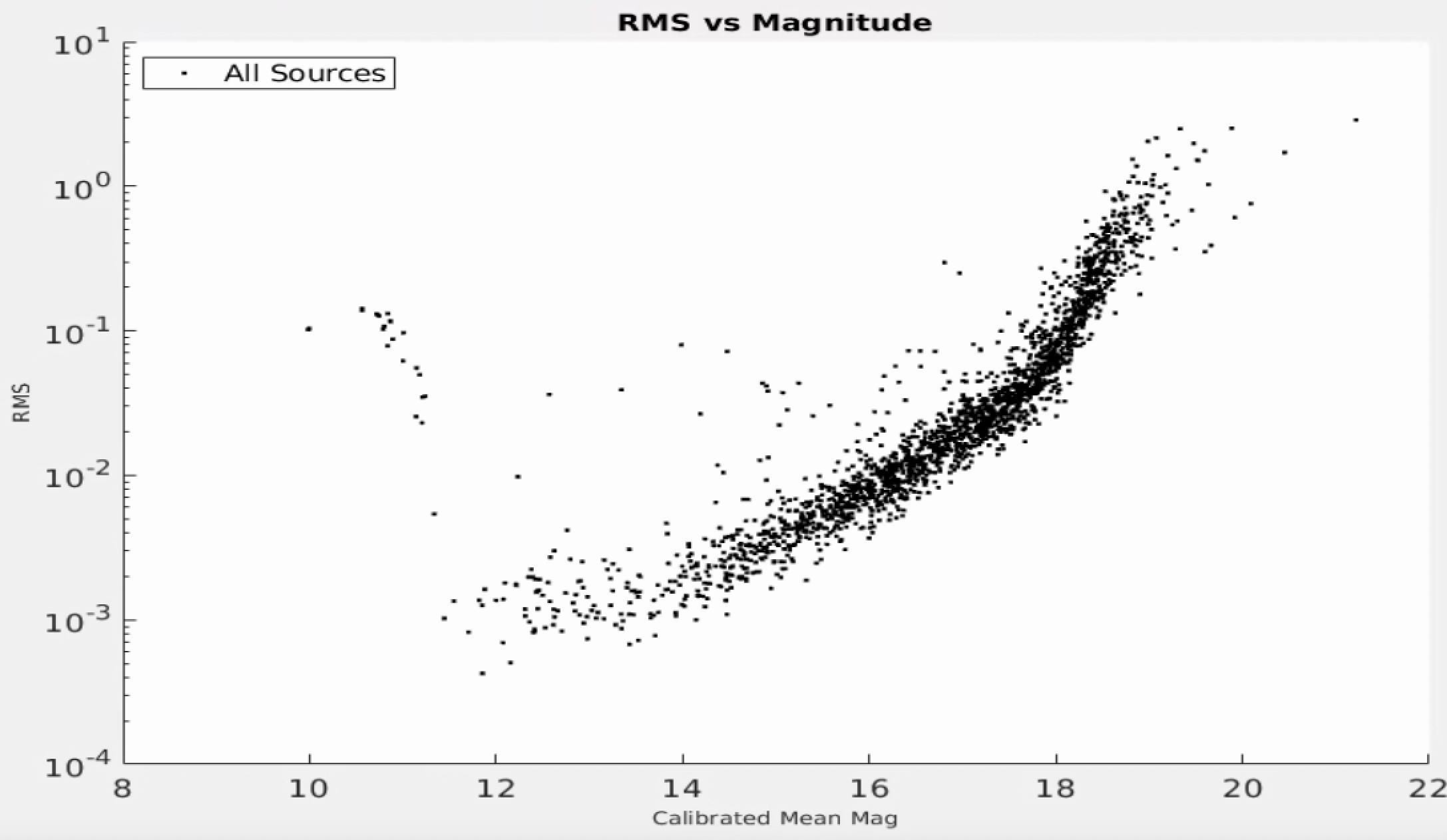}}
\caption{Root Mean Square (RMS) vs Magnitude} 
\label{}
\end{figure}

\section{Variability Report Generator}
\label{varrefgen}
In this section, we describe a novel variability report generator developed by the author using the {{\tt AstroPack}/{\tt MAATv2}} package in MATLAB. Designed to streamline the process of identifying variable sources in LAST's dataset, our custom-built code efficiently analyzes each source within a given sub-image, utilizing specific thresholds to identify candidate variable stars. Specifically, the report generator employs a periodogram cutoff, evaluating the power spectrum of each source to determine whether it surpasses a designated sigma value. Additionally, an RMS threshold is applied to detect sources with photometric variations further away from the median, thereby flagging them as potential (albeit non-periodic) variable stars. By incorporating these criteria, the generator automates the identification of variability in a LAST image, saving valuable time and effort in the analysis process. Upon identifying the candidate variable sources, the program automatically generates visual photometric reports for each source. These include a binned light curve, a folded light curve, a plotRMS and a periodogram depicting the dominant periods associated with each candidate variable source. In addition to the visual reports, the program compiles a photometric tabular report containing essential information for each candidate variable source, providing details such as period, amplitude, mean, RMS and other relevant properties. Such a report eases the process of identifying the few proverbial needles exhibiting variability in the haystack of candidate variables. In this section, we offer a walkthrough of the MATLAB code used for generating a variability report from a LAST sub-image. The script encompasses various steps, starting from calibration to detection to report generation.

\subsection*{Step 1: Creating Matched Sources Objects}

The script begins by creating two matched sources (MS) objects, namely \texttt{MS1} and \texttt{MSall}. \texttt{MS1} contains matched sources from a single visit, while \texttt{MSall} contains matched sources from all visits (24 subimages). The data are read from the specified directory path, and the user has the option to provide a custom path or use the default dataset.

\subsection*{Step 2: Matching Sources by Sub-Image}

Next, the script matches sources based on the sub-image number (\texttt{n}). It utilizes the \texttt{mergeByCoo} function to match sources from the \texttt{MSall} object to the corresponding \texttt{MS1} object for the specified sub-image.

\subsection*{Step 3: Zero-Point Calibration}

The script proceeds with zero-point calibration for the matched sources based on aperture or PSF-fit photometry. It calculates the calibration parameters (\texttt{R}) and applies the zero-point correction to create a new ZP-calibrated matched sources object named \texttt{MSr}. The \texttt{MAG\string_APER\string_3} field with a 6-pixel radius aperture is used for error calibration in both aperture and PSF-fit photometry.
  
\subsection*{Step 4: Binning the Data}

To group successive epochs together, the script bins the data using the \texttt{binning} function. The user can input a custom binning size or use the default value (10). The binned data are stored in the \texttt{MSr\string_binned} object.

\subsection*{Step 5: Generating Periodograms}

The script generates periodograms for all sources in the \texttt{MSr} object. It sets specified frequency limits and peak power thresholds to identify potential periodic variables. The user can customize the thresholds, or default values are used -- $(100, 1000)$ per day for the frequencies and $13 \sigma$ for the periodogram -- if not provided.

\subsection*{Step 6: Polynomial Subtraction}

Systematic effects are removed and variability is extracted by performing polynomial subtraction on both the unbinned and binned light curves. The script fits a third-degree polynomial to the data and subtracts it to obtain the subtracted magnitudes (\texttt{M\string_os} for unbinned and \texttt{M\string_bs} for binned).

\subsection*{Step 7: Plotting Periodograms and RMS}

Periodograms (power spectra) are plotted for all sources and for selected sources that surpass the peak power threshold. The power spectra are represented as ampltiude ($\sigma$) vs. frequency. Additionally, the script plots RMS (Root Mean Square) vs. magnitude for all sources, where RMS serves as a measure of variability.

\subsection*{Step 8: Selecting Sources}

The user is prompted to enter a source index or (RA, Dec) coordinates to select a specific source for detailed analysis. If left empty, the script automatically chooses the default source with the highest peak in the periodogram.

\subsection*{Step 9: Plotting Light Curves}

The script plots both the unbinned and binned light curves for the selected source. The unbinned light curve shows data points using \texttt{M\string_os} while the binned light curve displays the data points after binning (using \texttt{M\string_bs}).

\subsection*{Step 10: Folded Light Curve}

A folded light curve is generated for the selected source using the frequency of the peak in the periodogram. The script folds the data points according to the dominant period of the source's variability and plots the folded light curve, revealing the repeating pattern of its brightness fluctuations.

\subsection*{Step 11: Tabular Photometric Report}

Finally, the script generates a tabular photometric report for selected sources with information such as period, amplitude, mean magnitude, median magnitude, robust standard deviation, standard deviation of magnitude, and RMS. The report is displayed in a table format for easy analysis and comparison.

\subsection*{Improvement Areas}

The code author has identified certain areas for improvement as of August 2023. These include implementing polynomial subtraction on the power spectrum, setting an RMS threshold for identifying non-periodic variable stars, flagging problematic sources based on bit masks, including $\Delta\chi^2$ based on polynomial fit in the table report, addressing inconsistencies in binning methods, and debugging the window function for sampling periodicity.

\subsection*{Usage and Acknowledgment}

The code was authored by Barkotel Zemenu in August 2023 and is originally designed for analyzing data from LAST. Users can execute the script as demonstrated in the example below:

\begin{verbatim}
>> generatevariabilityreport
Enter sub-image number [1, 24]: 10
Enter the source index or (RA, Dec) coordinate:
(3.022022977394895e+02, 65.349260622926622)
\end{verbatim}

\subsection*{Output}
\begin{figure}[hbt!]
\centerline{\includegraphics[width=13cm]{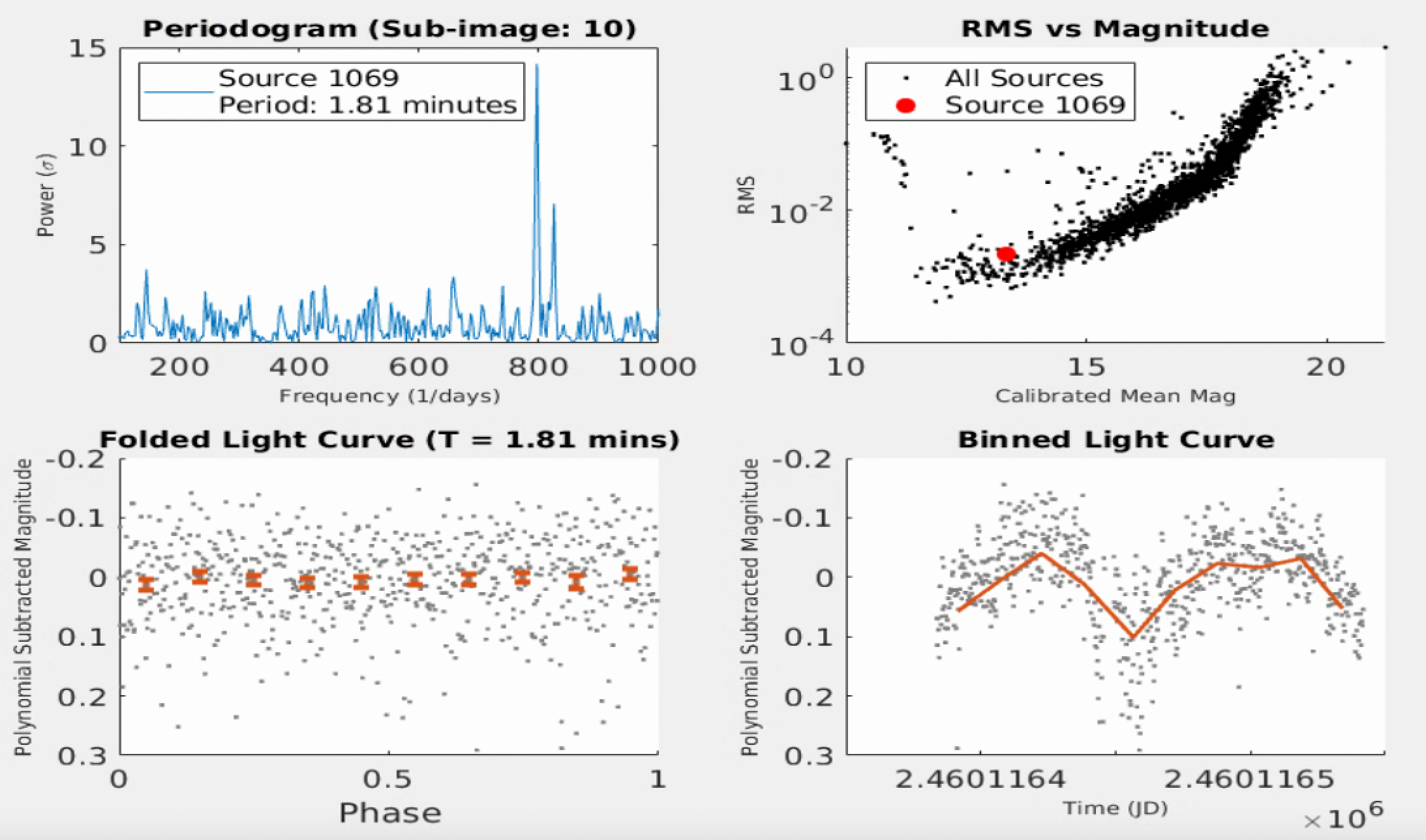}}
\caption{Visual Photometric Report for Candidate Variable Source} 
\label{output}
\end{figure}

The program has five input options (sub-frame number, bin value, minimum/maximum periodogram frequency, and peak power threshold), only one of which is not optional (sub-frame number). With just a single input, the program generates a visual report for a selected source and a tabular report for all the candidate variable sources (see Fig. \ref{output}).

\begin{figure}[hbt!]
\centerline{\includegraphics[width=13cm]{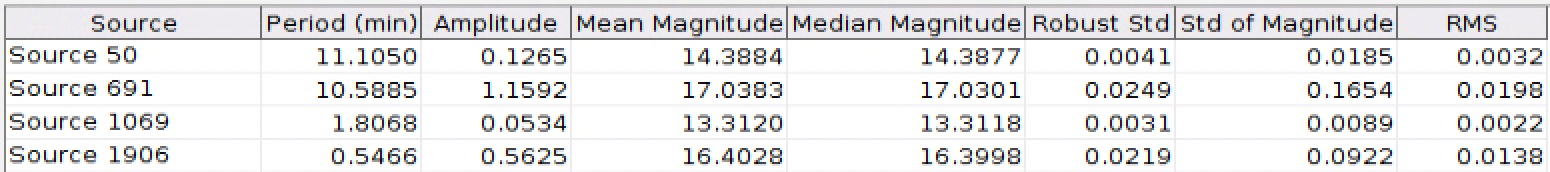}}
\caption{Tabular Photometric Report for Automatically Identified Candidate Variables} 
\label{}
\end{figure}

\newpage
\section{Conclusion}
\label{conclusion}
This paper provided a comprehensive look at the Large Array Survey Telescope (LAST)'s variability search performance. The technical aspects of the LAST observatory were reviewed, focusing on its imaging strategy and system design. The data reduction pipeline was described in detail, highlighting the calibration procedures that ensure accurate data for analysis. Variability search tools were explored, with emphasis on analyzing light curves, periodograms, and RMS plots. Finally, the variability report generator, a custom-built program aiding variability searches, was introduced. The program's efficiency and automation capabilities streamline the process of identifying and characterizing candidate variable sources, making it an indispensable tool for efficiently extracting variability insights from the vast amount of observational data produced by LAST.

\section*{Acknowledgments} The author expresses gratitude to the {LAST research team} in general and Prof. Eran Ofek, Dr. David Polishook and Yarin Meir Shani in particular. This research was conducted as part of the Kupcinet-Getz International Summer School at Weizmann Institute of Science in summer 2023. For official LAST publications, please consult \cite{sciencepaper}, \cite{overviewpaper}, \cite{pipelinepaper} and the \href{http://www.weizmann.ac.il/wao/large-area-survey-telescope-last}{Weizmann Astrophysical Observatory} official website.

\newpage
\appendix
\section{Matched Filtering}
\label{app1}
Matched-filtering is a signal processing technique we employ to improve the detection of particular signals present in LAST images. It involves convolving an image image with a template (or filter) that is designed to match the expected shape of the signal of interest. In LAST, the template used is a Gaussian filter. Below are some key concepts related to matched-filtering:

\textit{Gaussian Filter Template:} A Gaussian filter is characterized by its center position, width (standard deviation), and amplitude. The Gaussian filter is chosen as our template for its desirable properties, such as smoothing and preserving the spatial information of the image.

\textit{Convolution:} The next step is to convolve the image with the Gaussian filter template. Convolution is a mathematical operation that combines the pixel values of the image with the template in a way that emphasizes regions where the template closely matches the image. The discretized convolution operation can be represented by the equation:
\begin{equation}
    M'(\mathbf{x}) = \sum_{\mathbf{k}} T(\mathbf{k}) \cdot I(\mathbf{x} - \mathbf{k}),
\end{equation}
where $M'(\mathbf{x})$ is the matched-filter response image, $T(\mathbf{k})$ is the Gaussian filter template, and $I(\mathbf{x} - \mathbf{k})$ is the image being convolved with the template.

\textit{Template Matching:} During convolution, the Gaussian filter template is slid across the image, and at each position, the template is multiplied element-wise with the corresponding image pixels. The resulting values are summed to generate the matched-filter response image $M'(\mathbf{x})$.

\textit{Signal Detection:} The matched-filter response image highlights regions in the original image that closely resemble the shape and characteristics of the Gaussian filter template. It effectively enhances the presence of signals or features that match the template. Higher values in the matched-filter response image indicate a stronger correlation between the template and the corresponding regions in the original image.

\textit{Thresholding and Analysis:} Once the matched-filter response image is obtained, further analysis can be performed. This may include applying a threshold to identify significant detections or extracting quantitative measurements of the detected signals, such as their positions, sizes, or intensities.

The matched-filtering technique with a Gaussian filter template is often used for various purposes, such as detecting and characterizing point sources (e.g., stars), identifying extended structures (e.g., galaxies), or enhancing specific features in the data (e.g., faint signals or low contrast structures). Applying the matched-filtering technique with a Gaussian filter template effectively enhances the detectability of specific signals or features of interest, enabling more accurate analysis in LAST images.

\section{Photometry}
\label{app2}
By photometry, we are referring to the measurement of the brightness of sources in LAST images. The process involves quantifying the amount of light coming from celestial objects, such as stars, galaxies, or other astronomical sources, captured from all the visits. While photometry can be performed in several ways, two key methods are utilized in LAST: {aperture photometry} and {PSF-fit photometry.}

\subsection{Aperture photometry}
Aperture photometry involves measuring the total amount of light within a predefined circular -- and sometimes elliptical -- aperture centered on the object of interest. This method provides a simple and intuitive way to estimate the object's brightness, but it can be affected by background noise and contamination from nearby sources. To perform aperture photometry, we first select a circular or elliptical region around the target object in the astronomical image. The size of the aperture is typically chosen based on the expected size of the object and the level of background noise in the image. The aperture is centered on the object's position to ensure that we measure the majority of the light coming from the source. Next, we sum the pixel values within the aperture to obtain the total flux or brightness of the object. This sum includes both the light from the object itself and any background noise or contamination from nearby sources that falls within the aperture. To account for the background noise, we often subtract the average background value per pixel from the total sum. The resulting value is the aperture photometry measurement, representing the estimated brightness of the object. However, it is essential to consider factors like the aperture size and background estimation carefully, as they can significantly influence the accuracy and precision of the photometric measurement. Aperture photometry is particularly useful for bright and well-isolated sources in the image, where contamination from neighboring objects is minimal. It provides a quick and straightforward way to estimate the brightness of celestial objects and is commonly used for studies involving point sources, such as stars. However, despite it being relatively straightforward, aperture photometry has limitations. For extended or faint sources, the choice of aperture size can become challenging, as larger apertures may introduce more background noise, while smaller apertures may miss some of the light from the source. In such cases, PSF-fit photometry is often preferred as it offers a more sophisticated approach to handle varying source sizes and background conditions.

\subsection{PSF-fit photometry}
In real-life situations, when we observe sources (like stars or galaxies), there is something called a point spread function (PSF) that affects how they appear in our images. This means that different pixels in the image containing the source will have different amounts of light, or photon counts. Because of this, we can assign different {weights} to different pixels to account for these variations. A more naïve -- but sometimes justified\footnote{If the source noise dominates over the background, readout and/or dark current noise, for instance.} -- approach would be to equally weigh all pixels assuming that the source uniformly. One important aspect we need to consider, however, is the signal-to-noise ratio (SNR), which tells us how strong the signal (the source we want to detect) is compared to the background noise. Our ultimate goal, of course, is to choose a weight that maximizes the SNR. In the case of background-dominated Gaussian noise, the best weights to use (read: those reducing source detection uncertainty) are the ones that match the shape of the PSF. This means that we give more importance to pixels that are closer to the center of the source. Of course, this is assuming we know a source's PSF in the first place, which is not often true.\footnote{For a Gaussian PSF -- which approximates the PSF (Airy Function) quite well -- the maximal SNR can actually be determined in closed form.} On the other hand, PSF photometry takes into account the shape of the point spread function, which describes how light from a point source spreads out in the image due to the effects of optics and atmosphere. The PSF is typically modeled as a Gaussian or a more complex function estimated based on bright sources. PSF photometry involves fitting the PSF model to the observed data around the object and extracting the flux, position and even background from the best-fit model. This technique provides a more accurate estimation of the object's brightness, especially in crowded fields where there are faint sources. 

{Convolution-based PSF photometry} is a particular type of PSF photometry employed in LAST. Think of convolution as a way to combine the information in the image with the information in the PSF. We start with the image we captured, which consists of pixels that represent different amounts of light. Then, we take the PSF, which describes how the light spreads, and we apply it to every pixel in the image. The PSF is like a "template" that we slide over each pixel in the image. As it slides, it combines the light from that pixel with the light from its neighboring pixels, just like how the actual spreading of light happens. After going through every pixel, we end up with a new image where the light has been "convolved" or blended according to the PSF. This new image gives us a better understanding of how the light from the object is distributed. Now, to measure the amount of light coming from the object, we can examine the convolved image. We look at a specific spot in the image that corresponds to the object's position. By reading the value of the pixel at that spot, we can estimate the amount of light coming from the object. 
In sum, convolution helps us account for the spreading of light in the image by combining the information from the image with the PSF to create a convolved image. By analyzing the convolved image, we can estimate the amount of light coming from sources in a LAST image.   

\printbibliography

\clearpage

\printglossary[title={Glossary},toctitle={Glossary}]

\end{document}